# Sharp, localized phase transitions in single neuronal cells

Carina S. Fedosejevs[1]*, Matthias F. Schneider[1]*

[1]Department of Medical and Biological Physics, Technical University Dortmund; 44227 Dortmund, Germany.

*Corresponding authors. Carina S. Fedosejevs: Otto-Hahn-Straße 4, 44227 Dortmund, Germany, +49 231 755 2994, carina.fedosejevs@tu-dortmund.de.
Matthias F. Schneider: Otto-Hahn-Straße 4, 44227 Dortmund, Germany, +49 231 755 4139, matthias-f.schneider@tu-dortmund.de.

**Email:** carina.fedosejevs@tu-dortmund.de (C.S.F.); matthias-f.schneider@tu-dortmund.de (M.F.S.)

**Abstract**

The origin of nonlinear responses in cells has been suggested to be crucial for various cell functions including the propagation of the nervous impulse. In physics nonlinear behavior often originates from phase transitions. Evidence for such transitions on the single cell level, however, has so far not been provided leaving the field unattended by the biological community. Here we demonstrate that single cells of a human neuronal cell line, display all optical features of a sharp, highly nonlinear phase transition within their membrane. The transition is reversible and does not origin from protein denaturation. Triggered by temperature and modified by pH here, a thermodynamic approach, strongly suggests, that similar nonlinear state changes can be induced by other variables such as calcium or mechanical stress. At least in lipid membranes such state changes are accompanied by significant changes in permeability, enzyme activity, elastic and electrical properties.

**Significance Statement**

Phase transitions in materials are accompanied by drastic changes in their properties. Systems abruptly become softer, more conductive, a better heat storage or support chemical reactions more efficiently. Since changes take place over small variations in external conditions (tension, temperature, pH, calcium..) they appear like an on/off switch. Here, we provide experimental evidence that membrane patches of single living cells can go through a reversible phase transition. It is extremely "sharp" (highly nonlinear) and from a thermodynamic point of view we conclude, it cannot only be triggered by temperature, but also by pH changes (as produced by enzymes). The results strongly support the idea that phase transitions may be a tool for living systems to control their functions even specifically.

**Main Text**

**Introduction**

Phase transitions are not only an attractive field for solid state, but also for soft matter and biophysicists. During a phase transition, material properties change dramatically upon applying a fairly moderate or small external stimulus. Both, synthetic and native membranes have been provided an attractive target to explore the existence of transitions in living organisms and their impacts on cell functions such as membrane transport, enzyme activity and mechanical properties (1–7). Their existence has been already demonstrated within cell ensembles, but it has often been claimed that these transitions are too smooth to be biologically relevant (1, 2). The goal here is to study transitions and their properties on a single cell level to provide insights into their significance for living systems. We use high resolution fluorescence microscopy to determine spectral changes upon temperature variation as a stable determinant for thermodynamic state changes.

**Results**

**Fluorescence as a reporter of local membrane state.** In Fig. 1 fluorescence images of a giant unilamellar lipid vesicle (Fig. 1A) and living cells from the *SH-SY5Y* cell line (Fig. 1B and C) are shown. The lipid anchored dye *Atto488 dipalmitoylphosphatidylethanolamine* (*DPPE*) is directly incorporated into the lipid or cell membrane and acts as a state-probe, i.e. it reports changes in membrane state due to the states impact on the emission properties. This has been demonstrated a powerful technique to detect phase transitions in lipid membranes (8–11). *Atto488 DPPE* is highly photostable and was characterized to respond, with a jump in the emission spectrum of ~10 nm, to a state change in the surrounding lipid membrane (Fig. 1D). This behavior is measurable for zwitterionic dimyristoylphosphatidylcholine (DMPC) lipids as well as for charged lipids such as dimyristoylphosphatidylserine (DMPS). Although the spectral shift of the fluorescence emission spectrum of *Atto488 DPPE* is rather small (compared for instance with Laurdan (8, 12) or Di-4-ANEPPDHQ (10)) it is clearly resolvable in a confocal system when combining narrow detection filters and highly sensitive detectors. We record the ratio at 520 nm and 530 nm (the intensity ratio parameter $r$) with a width of 10 nm each, representing two flanks of the spectrum (Fig. 1D) as a measure for the spectral shift. The dye has a lipid anchor (DPPE) and therefore a high affinity for embedding itself into the cell membrane. A cross-sectional scan of a single cell (Fig. 1C) demonstrates that the dye is not diluted in the media but associated to the cell membrane. Within the confocal setup it is possible to excite a small patch of only ~ 1 µm$^2$ in the two-dimensional cellular membrane and monitor its response locally upon temperature variation (Fig. 2). However, the signal within the confocal volume must be considered a superposition with the background signal due to the size of the confocal volume (~ 0.8 µm) compared to the membrane thickness (~ 10 nm). Free and membrane bound dye can be separated based on their diffusion constants, since diffusion times increase by an order of magnitude within the membrane (13) (Fig. S1). In order to reduce background noise from dye which was endocytosed into vesicles we avoided areas with larger accumulations of fluorescence signal within the cell. Finally, pure dye does not interfere significantly with the signal from the membrane, since it does not represent any significant temperature dependence around 10 °C – 50 °C (Fig. S2).

**Phase transitions in single cells.** Our key finding is shown in Fig. 3 and demonstrates the existence of a sharp phase transition within a ~ 1µm$^2$ membrane patch of a single living cell. The intensity ratio $r$ is plotted as a function of temperature for the patch on the living cell (Fig. 3B) and directly compared to $r$ within a suspension of lipid vesicles of a well-known studied phase transition (Fig. 3A). A sudden jump, indicating an abrupt shift in emission frequency spectra, is shown in both systems with the same sign and order of magnitude. Note, absolute values of the ratio differ due to differences in the initial position of the emission spectrum (compare Fig. 1D). Since lipid vesicles of DMPS are known to go through a solid ordered – liquid disordered transition around 38 °C (14), the jump in spectra detected within a single spot on the cell membrane of these neuroblastoma cells clearly demonstrates that the cell membrane goes through a phase transition.

To quantify the results the temperature with the greatest change in ratio or rather the maximum of the derivative of that sigmoidal-like curve is used to assign an explicit phase transition temperature ($T_m$). Note, an ongoing change in the spectrum shape, as visible in Fig. 1D, causes an additional corresponding drift in the ratio over the whole temperature range (Fig. 3A). The full width at half maximum (FWHM) of the derivative is a measure for the width of the transition. With this, $T_m$ = (40.4 ± 0.1) °C and FWHM = 2.2 °C is derived for the measurement on the DMPS vesicle suspension in Fig. 3A which is in good agreement with the literature (14) under the given measurement conditions (see methods). Further, within the membrane patch of the living cell measured in Fig. 3B the phase transition temperature $T_m$ is located at (17.1 ± 0.1) °C and the width is calculated to FWHM = 1.0°C. Overall, the transition was measured on more than 10 different single cells (at one randomly picked spot each) and the mean $T_m$ over all these measurements at pH 7.4 (n=13) is (18.2 ± 0.4) °C which is around 20 °C below the growth temperature of the cells (37 °C). The average FWHM is (0.96 ± 0.13) °C whereas the average FWHM within the vesicle suspensions (n=8) is (2.66 ± 0.22) °C.

**Highly nonlinear transition**. Importantly, the transition in single cells is highly nonlinear as it is remarkably sharp with a FWHM of only 1 °C or less (Fig. 3B). This presented nonlinear behavior was repeatedly observed in measurements across different single cells of the *SH-SY5Y* line whether the cells were cooled or heated in the process. The transition temperatures varied throughout the measurements mainly between 15 °C and 20 °C and the transitions were consistently sharp [$\overline{\text{FWHM}}$ = (0.96 ± 0.13) °C]. The general existence of phase transitions in cells and especially in mammalian cells has been often observed in calorimetric (2, 15) and optical measurements (4, 6, 16, 17). Transition temperatures in cells were found typically between 10 °C to 30 °C below growth/living temperature, which is in good agreement with our observation. However, transitions in such ensemble measurements are consistently considerably broader (15, 16, 18) than what we are reporting here on single cells. We believe this is a consequence of ensemble averages and will discuss this point further below.

**Transitions are reversible.** Figure 4 demonstrates that cells can repeatedly and reversible go through the transition. This further corroborates the idea that the transition is within the lipid matrix of the membrane as transitions in proteins would be expected to be rather irreversible (2). A strong correlation of $T_m$ measured in whole cells and the $T_m$ of their extracted membrane lipids could be shown in calorimetric measurements already (2, 15) hinting the significance of the lipids in this process. Throughout the cooling and heating cycle a path dependency is visible in agreement with hysteresis effects well known from the theory of phase transitions (19). During heating the transition tends to be broader, meaning it takes more time for the system to accommodate to the temperature changes in order to break up the higher ordered structure of the gel-like phase. A less pronounced "step" in the heating process following the cooling process is most likely due to a smaller amount of dye participating in the process. Already mentioned metabolization effects such as endocytosis continue over time and lower the amount of dye in the membrane constantly. The background signal is expected to increase and the signal to noise ratio gets worse over time.

**Why is the transition in single cells so sharp?** Phase transitions studied on large samples using for instance calorimetry show a very broad transition of ~ 10°C (2, 15). The large width of the transition has been put forward in order to disregard the role of phase transitions in biology. Here, however, we demonstrate, that the transition within a single cell membrane is just as sharp or even sharper than the one in pure synthetic, one component lipid vesicles and can take place over less than 0.6° C (compare Fig. 4). We have earlier suggested, however, that the wide range of the transition region in calorimetry is simply a result of the large sample sizes used (20). One of the most intensively studied system is for instance the bacteria E. Coli., where at least millions of bacteria within the calorimeter cell are responsible for the final signal (15). We believe that there are at least two sources, which widen the heat capacity signal in such experiments: heterogeneity within and across the cells. First, each cell consists of a lot of different materials in particular the intracellular protein networks and/or the cell membrane. Further, no two bacteria will have exactly identical transition points $T_m$, as they don't have identical compositions. Just like in Eigen's theory of the quasispecies, we would expect a distribution of individual $T_m$'s scattered around some "center of gravity" (21). Integrating over all those $T_m$'s results in a broadening of the heat capacity profile. In fact, we observe a spreading in $T_m$ mainly between 15 °C and 20 °C throughout the experiments. The width is constantly between 0.4 °C and 1.6 °C [$\overline{\text{FWHM}}$ = (0.96 ± 0.13) °C]. In direct comparison these transitions tend to be sharper than those measured in lipid vesicles [$\overline{\text{FWHM}}$ = (2.66 ± 0.22) °C]. However, in cells we measure localized within a small membrane patch (~ 1 µm$^2$) whereas in a suspension of lipid vesicles a certain order of heterogeneity is introduced due to distributed radii and curvature of the vesicles (2, 9, 22).

**Sensitivity.** Although quantitative numbers of mechanical or thermal properties of membranes cannot directly or easily be extracted from optical measurements, our approach suits small and alive samples and is highly sensitive. The order of magnitude of membrane lipids needed in modern bio-calorimetry is in the order of $10^{-5}$ g - $10^{-4}$ g if the sample is very pure (e.g. synthetic lipids). Here, we analyze the thermodynamic behavior of a small membrane segment of an individual cell. The area probed is ~ 1µm$^2$ which corresponds to ~ $10^6$ lipids or a weight of ~ $10^{-15}$g or 1 femtogram, which is ~10 orders of magnitude less at least (since we actually only observe the dye which is 0.4 mol% or less). I.e. our approach allows to study phase transitions localized on living cells in small and confined space.

**Coupling of temperature and pH.** Our study is focused on transitions induced by temperature so far as it is an experimentally convenient accessible parameter where no intervention in the system is required during the measurement (e.g. exchange of cell buffer solution). Further, it is the most intuitive and well studied parameter in the context of phase transitions. However, it is well established - both theoretically and experimentally - that thermodynamic variables (pressure, electrical field..) are in general coupled and affect the phase state as well (15, 23). A parameter of particular physiological relevance is the pH and theoretical predictions have been established by Träuble concerning the existence of the coupling of temperature and pH at the membrane as well (24). Due to (pK-dependent) acidic lipid headgroups, the membrane is receptive to protons, which thereby control the surface charge $\sigma$. Träuble investigated the electrostatic effects on the membrane structure with the Gouy-Chapman theory and derived an equation (24) where the impact of the pH and salt concentration *n* on the phase transition is included:

$$\Delta T_m = -2 \frac{kT}{e} \frac{L}{\triangle S^*} \sigma \Delta f + \frac{\varepsilon}{\pi} \left(\frac{kT}{e}\right)^2 \frac{L}{\triangle S^*} \kappa \Delta f$$

Here $\Delta T_m$ represents the resulting shift in the melting or transition temperature, $\Delta f$ are the changes in molecular area of the lipid molecules L between the two phase states (e.g. ordered and disordered), $\triangle S^*$ is the change in entropy during the transition and $k, e$ and $\varepsilon$ are well known constants (Boltzmann constant, elementary charge and dielectric constant). While the 2nd term in the Eq. accounts for the impact of ions (salt concentration $n \sim \kappa^2$ where $\kappa$ is the Debye length) on the melting temperature, the first term represents the role of the pH since pH strongly influences the surface charge $\sigma$. Upon protonation (by lowering the pH) the negatively surface charge of the (cell) membrane is lowered, which reduces lateral, repulsive forces. As a result, the membrane is more condensed at otherwise constant conditions, which in turn increases the transition/melting temperature $T_m$. To push this thought further, it would be even possible to induce a phase transition merely upon protonation at a certain (constant) temperature. This behavior is both, correctly predicted by Träuble's equation as well as experimentally verified (24). Additionally, the relation between $T_m$ and pH has been studied in many artificial membrane systems as well as cells, where the negatively charged membrane is receptive to protons (1, 4, 25). A thorough review has recently been published by our group (18). Note, the membrane is most sensitive to pH changes in the vicinity of its pK, where a large degree of condensation is expected upon small changes in pH. Further away from the pK the response to changes in pH is less pronounced. Owing to this range in susceptibility, a decrease in one pH unit may result in an increase in $T_m$ between 2°C – 35°C (refer to Table 2 in (18)) depending on the system and its state.

To test our results for the general coupling we lowered the pH from 7.4 to 5.5 and varied the temperature from 35 °C to 21 °C (Fig.3C). In very good agreement with the discussed theory and experiments, we find that the transition temperature is increased by ∼ 13 °C compared to $\overline{T_m}$ at pH 7.4 for the given example in Fig.3C. Overall, we get $\overline{T_m}$(pH 5.5) = (29.9 ± 0.8) °C (n = 4) and with that an increase by ∼ 12 °C compared to $\overline{T_m}$(pH 7.4) corresponding to an increase of ∼ 6 °C per pH unit. Note, this derived factor for the coupling is not expected to be linear over a wide pH range as it was discussed in the paragraph before. Based on our data, the width of the transition is not affected by pH as it remains with $\overline{\text{FWHM}}$ = (1.05 ± 0.19) °C within the order of magnitude of the width at pH 7.4 [$\overline{\text{FWHM}}$ = (0.96 ± 0.13) °C]. However, this demonstrates clearly that the pH is a potent regulator of the phase state of the cell membrane.

**Discussion**

Our results demonstrate on a single cellular level, that membranes of living cells can go through sharp, reversible transitions. This opens up a vast amount of ideas that date back at least to the 70ties and the work of Träuble as well as Changeux among others who have suggested that phase transitions may play an important role in biology (2, 4, 6, 7, 26, 27). Sharp, nonlinear behavior in material properties may provide an origin of many switch-like functions (1, 26). If a lipid membrane for instance suddenly softens by an order of magnitude, chances for morphological transitions as they occur during adhesion, budding, fission and fusion are tremendously increased (20, 28).

Even ∼ 20 °C below physiological temperatures the transition is relevant. It is important to recall, that applying temperature to invoke a transition, as it has been done here, is a mere matter of experimental convenience. From a thermodynamic point of view it is strongly suggested that physical parameters such as temperature, pressure as well as chemical parameters, e.g. pH or calcium concentration etc. are able to trigger transitions (1, 2, 18). For instance, it has been shown for monocomponent lipid systems (1) and predicted for cells (18), that it is possible to shift $T_m$ and compensate 10°C - 20°C by 1-2 pH units. Our results fit into this range quite well and confirm the predicted role of the pH. The shift in $T_m$ upon a shift in pH is, thermodynamically, equivalent to the possibility to induce a phase transition by pH at constant temperature. Thus, just noise will not be enough to introduce a transition, which would render the system too sensitive to external perturbations. A local change of ~1 pH unit, however, might be enough to shift the system very close to its transition regime from where only a slight variation/perturbation would trigger the transition due to its sharpness. The local pH concentration required clearly depends on the local lipid composition of the membrane and in terms of the pH, especially the headgroups of the lipids and the local pK of the membrane (1). The latter would be a crucial parameter in this picture and "decides" whether a locally invoked pH shift, introduced for instance by the activity of an enzyme, would be sufficient to alter the state of the system. Acetylcholinesterase for instance, can reduce the local pH from 7 to 4.5 within less than 1ms (25). This introduces a specificity for the ability to undergo a phase transition under certain conditions or in other words some kind of excitability.

Excitability or switching of membrane-controlled functions by altering the phase state have been discussed and observed earlier (1, 3, 4, 6, 10, 29) and was strongly supported by the work on lipid membranes. In the latter, one was able to demonstrate a state-function relation in several parameters, such as permeability, budding, fission, adhesion and catalytic activity (20, 28, 30). The sharpness of the phase transitions found here in single living cells supports the idea, that state changes are promising candidates to adopt membrane function in a highly nonlinear (switch-like) matter. Whatever is capable to change state, is in principle a possible trigger for a transition. Electrostatically, calcium or any (in particular higher charged) ion could be potent state regulators and mechanically, applied tension would alter state and hence function (1, 18, 23). Finally, molecules (drugs or proteins) which incorporate within the membrane should affect the state as well.

The response of the system to such triggers are alterations (transitions) in membrane properties. There is strong evidence in literature that they can not only occur in thermal, mechanical or electrical parameters but also in catalytic activity or conductivity (30, 31). Our results are especially interesting since we use neuronal-like cells which are well known for excitability regarding their signal transducing purposes where nonlinearities and threshold-like behavior are present characteristics as well (32).

Our findings offer a very new avenue to think about the physical origin of nonlinear regulation in biology in general and further support the idea, that the nonlinear behavior of action potentials is tightly linked to the nonlinear material behavior of the cell membrane reported here (3, 5, 10, 29).

**Materials and Methods**

All chemicals were purchased from Sigma Aldrich with a purity of > 99% unless otherwise stated.

**Cell line.** *SH-SY5Y* cells (DSMZ, ACC 209) were cultured at 37 °C and 5% CO2 in medium containing Dulbecco´s MEM (PAN-Biotech) with 11% Fetal Bovine Serum (FBS) (PAN-Biotech) and 1% Penicillin-Streptomycin solution (PAN-Biotech). For sample preparation cells were transferred and grown on a cleaned coverslip.

**Staining of the cells.** Prior to staining, cells are washed three times with buffer solution (5 mM HEPES, 2 mM $CaCl_2$, 3 mM KCl, 147 mM NaCl, pH 7.4). For staining 2 µl of a 0.5 mM stock solution of *Atto488 DPPE* (ATTO-TEC) in dimethyl sulfoxide (99.8%, Acros organics) are diluted into 2 ml of the buffer solution with the final concentration of 0.5 µM. Cells are incubated in the staining solution for 15 minutes at 37 °C and afterwards rinsed again for three times. For measurements at pH 5.5 the buffer was replaced (during staining and measurement) with a similar buffer which was titrated to pH 5.5 with HCl.

**Vesicle preparation.** For lipid vesicle production, first, lipids along with 0.1 mol% *Atto488 DPPE* were dried by evaporation of the storage solution (DMPC: chloroform, DMPS: 65/35/8 chloroform/methanol/water and *Atto488 DPPE*: 80/20 chloroform/methanol) under a flow of nitrogen and subsequent incubation in a desiccator for ~12 hours. Ultrapure water (arium pro, Satorius) is added to adjust a concentration of 2 mM lipids in the final suspension. The latter is heated 10°C above the respective $T_m$ (melting temperature) of the specific lipid type and repeatedly vortexed to form vesicles.

**Spectrometer measurements.** Fluorescence emission spectra were recorded with a spectrometer (lipid vesicles: Thorlabs Compact Spectrometer (CCS100/M); cells: Wasatch (WP-00183)) plugged into an Olympus IX71 microscope via an optical fiber. Samples were excited with a mercury burner.

**Fluorescence detection/Setup.** Fluorescence measurements were performed on a Microtime 200 device (PicoQuant) with an integrated laser line at 488 nm in pulsed mode (20 MHz), along with an Olympus IX73 microscope and a detection unit containing two single photon avalanche diodes (SPCM-AQRH-14-TR, Excelitas). To detect signal from different flanks of the spectrum two bandpass filters at 520 nm (10 nm) and 530 nm (10 nm) were used in combination with a high NA (1.2), 60X water immersion objective (UPLSAPO60XW, Olympus). Instead of water an oil substitute with similar spectral properties as water is used (Immersol W2010, Zeiss) to avoid evaporation of the immersion fluid during the measurement cycle. For cell as well as vesicle measurements a chamber (FCS2, Bioptechs) is used which is closable by a coverslip containing the sample. The coverslip as part of the chamber is coupled to the immersion objective. The chamber is connected to a heating bath (Lauda Eco Silver). Outer pipes ensure a flow around the sample with heating bath fluids and allow a temperature control of the sample. Temperature is measured with a thermometer (TMD-56, AMPROBE) coupled to the chamber with an accuracy of 0.1 °C. However, absolute temperature values may be affected by the coupling of the chamber to the non-heated immersion objective depending on the extent of the deviation between sample and room temperature.

**Measurement procedure.** At first, a cross-sectional scan is recorded to determine the position of the cell membrane (maximum of intensity) of a single cell. The heating bath is programmed to vary the temperature continuously between two fixed temperatures, with a gradient of ~ 0.3 °C/min. Every 1-2 minutes a data point is measured over 15s. Low laser intensities (10 µW) were chosen to ensure the viability of the cells (33–35). In between, cross-sectional scans were repeatedly recorded to assure the location of the measurement spot. Throughout the measurements (at pH 7.4 as well as at pH 5.5) viability of the cells was controlled by checking for morphological changes such as dislocation from the support.

**Data Analysis.** The software *SymphoTime 64* (PicoQuant) was used for data acquisition. For each data point intensity traces, time correlated single photon counting (TCSPC) histograms and autocorrelation curves for both wavelength channels were recorded simultaneously. Signal and background counts are distinguished and allocated based on the distribution of counts in the TCSPC histogram. The data is further processed with python scripts where, among others, ratios are calculated without background. The error of the ratio follows from the count rate error. A sigmoid is fitted to the data to meet the stepwise characteristic of the ratios as function of temperature and the transition temperature $T_m$ is associated with the maximum of its derivative. The width of the transition is measured by the full width at half maximum (FWHM) of the latter.

For the calculation of diffusion times the autocorrelation curves were fitted by applying standard models regarding one respectively two diffusing species in the confocal volume. For further information refer to publications of Bacia (13) and Eggeling (36).

**Figures**

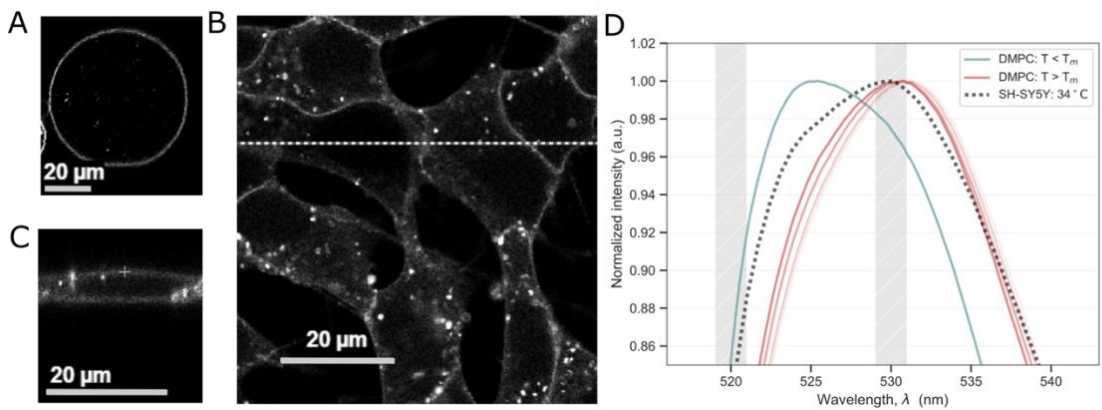

**Fig. 1: Confocal scans and fluorescence emission spectra of *Atto488 DPPE* incorporated into pure lipid vesicles and *SH-SY5Y* cells.** Brightness relates to intensity. (**A**) Giant unilamellar vesicle of phospholipids with *Atto488 DPPE* illustrating the basic principle of lipid anchored dye incorporation into a lipid membrane. (**B**) Top view scan of *SH-SY5Y* cells. The dye is embedded predominantly in the cell membrane. Bright dots of accumulated dye develop with progression of time due to uptake into the cell by endocytosis. The horizontal dotted line marks the position of the scan in C. (**C**) Cutout of a cross-sectional scan indicated in B, showing a single cell. The cross marks an exemplary measurement spot. (**D**) Normalized spectra in DMPC lipid vesicles (solid lines) from left to right with increasing temperatures and melting temperature $T_m$ = 24 °C (blue: T < $T_m$; red: T > $T_m$ with increasing translucence). The emission spectrum exhibits a jump around $T_m$ and a further, comparable small, linear shift during temperature variation. The dotted spectrum originates from a *SH-SY5Y* cell layer (~ 500.000 cells) at 34 °C. The hatched, grey areas indicate the location of the two wavelengths measured in the confocal setup from which the ratio parameter *r* is calculated.

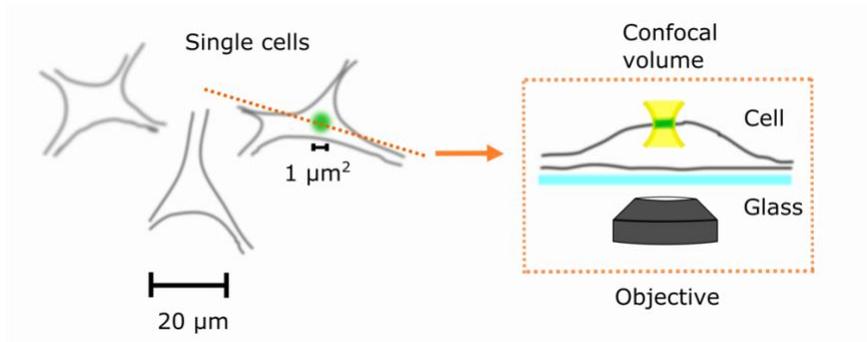

**Fig. 2: Fluorescence measurement in the dye labeled membrane of a single neuronal cell.** Two-dimensional membrane patch (~ 1 µm$^2$) inside the confocal volume is excited, on the left in the top view visible as green spot on a single cell and in the cross section on the right as green line within the confocal volume. Response from the lipid anchored dye molecules of this specific localized spot is monitored upon temperature variation at two wavelengths to extract a ratio parameter.

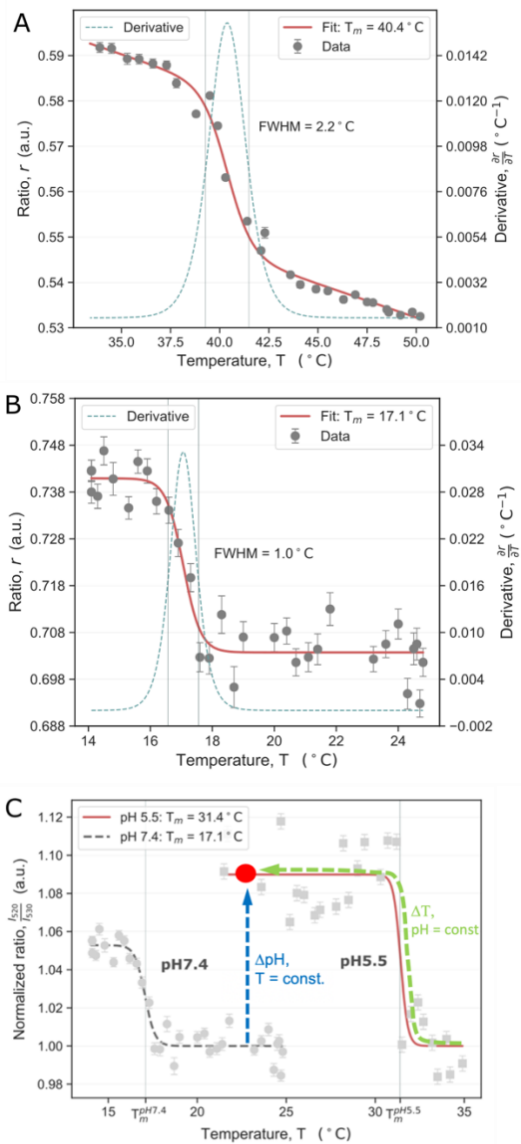

**Fig. 3: State changes from fluorescence spectra.** Ratio parameter $r$ ($I_{520\,nm}/I_{530\,nm}$) as a function of temperature. Each data point represents one measurement over 15 s and its errorbar results from the error propagation of the count rate uncertainty. **(A)** In a DMPS vesicle suspension $r$ (grey circles) displays a nonlinear and sharp jump indicating a spectral shift. The transition temperature extracted from the maximum of the derivative (dotted blue line) of the sigmoidal fit (red line) at $T_m$ = (40.4 ± 0.1) °C is in good agreement with the phase transition temperature of DMPS (38 °C (14)). The width is measured by the full width at half maximum (FWHM) of the derivative (FWHM = 2.2 °C). **(B)** Exemplary measurement at one patch in the cell membrane of a single *SH-SY5Y* cell. A nonlinear shift is detected at $T_m$ = (17.1 ± 0.1) °C [FWHM = 1.0 °C], resembling the shift in lipid vesicles (A) concerning sign and order of magnitude. Note, absolute values of the ratio differ due to differences in the initial position of the emission spectrum. These data reflect examples out of several measurement series, namely in 8 vesicle suspensions [$\overline{\text{FWHM}}$ = (2.66 ± 0.22) °C] and in 13 single cells [$\overline{\text{FWHM}}$ = (0.96 ± 0.13) °C]. **(C)** Coupling of $T_m$ and pH: Comparison of exemplary measured transition temperatures at pH 7.4 (grey dotted) and pH 5.5 (red line) demonstrating the general coupling between these two quantities. Here, $T_m$ is shifted from 17.1 °C to 31.4 °C upon acidification of the membrane by ~ 2 pH units. In total, at pH 5.5 (n=4) $\overline{T_m} = (29.9 \pm 0.8)$ °C was derived whereas $\overline{T_m} = (18.2 \pm 0.4)$ °C at pH 7.4 (n=13). This an approximate shift of 6 °C per pH unit which is within the theoretically predicted range (18). Bold dashed green line: Shift in state due to a change in temperature at constant pH. Bold dashed blue line: Shift in state due to a change in pH at constant temperature.

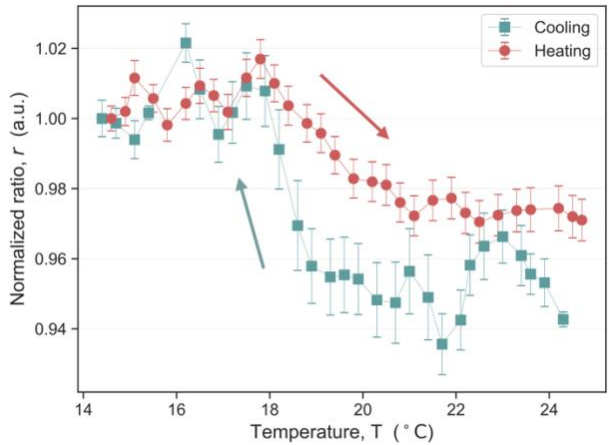

**Fig. 4: Reversible shift of the emission spectrum during cooling and heating.** Normalized ratio parameter $r$ ($I_{520\,nm}/I_{530\,nm}$) measured in the cell membrane of a single *SH-SY5Y* cell during cooling (blue squares) and subsequent heating (red circles) between 25 °C and 14 °C. Each data point represents one measurement over 15 s and its errorbar results from the error propagation of the count rate uncertainty. The beforehand described nonlinear shift in ratio is detected in both directions with $T_m = (18.6 \pm 0.1)$ °C [FWHM = 0.5 °C] for cooling and $T_m = (19.3 \pm 0.2)$ °C [FWHM = 1.6 °C] for heating (fit not shown for visibility).